\documentclass[twocolumn,A4]{article}
\usepackage[dvips]{graphics}
\begin{document}
\onecolumn
\begin{center}
{\bf{\Large On the role of electron correlation and disorder
on persistent currents in isolated one-dimensional rings}}\\
~\\
Santanu K. Maiti$^*$, J. Chowdhury and S. N. Karmakar\\
~\\
{\em Saha Institute of Nuclear Physics, 1/AF, Bidhannagar,
Kolkata 700 064, India}\\
~\\
{\bf Abstract}
\end{center}

 To understand the role of electron correlation and disorder on
persistent currents in isolated $1$D rings threaded by magnetic
flux $\phi$, we study the behavior of persistent currents in aperiodic
and ordered binary alloy rings. These systems may be regarded as
disordered systems with well-defined long-range order so that we do
not have to perform any configuration averaging of the physical
quantities. We see that in the absence of interaction, disorder suppresses
persistent currents by orders of magnitude and also removes its
discontinuity as a function of $\phi$. As we introduce electron
correlation, we get enhancement of the currents in certain
disordered rings. Quite interestingly we observe that in some cases, electron 
correlation produces kink-like structures in the persistent current as a
function of $\phi$. This may be considered as anomalous
Aharonov-Bohm oscillations of the persistent current and recent
experimental observations support such oscillations. We find that the 
persistent current converges with the size of the rings.

\vskip 1cm
\begin{flushleft}
{\bf PACS No.}: 73.23.Ra, 73.23.-b, 73.63.-b, 73.63.Nm\\
~\\
{\bf Keywords}: Persistent Current, Incommensurate Potentials, and Hubbard
Correlation.
\end{flushleft}
\vskip 2.5in
\noindent
{\bf ~$^*$Corresponding Author}:

Email address:  santanu.maiti@saha.ac.in
\newpage
\twocolumn

\section{Introduction}
In the mesoscopic and nanoscopic world phase coherence of the electronic 
states are of fundamental importance, and the phenomenon of persistent
current is a spectacular consequence of quantum phase coherence in these
regimes. In a pioneering work  B\"{u}ttiker, Imry, and 
Landauer~\cite{butt} suggested that even in the presence of impurity, a 
small conducting isolated ring enclosing a magnetic flux $\phi$
carries a current in the ground state, a current which {\em persists}
(does not decay) in time, and periodic in $\phi$ with periodicity 
$\phi_0=ch/e$, the elementary flux quantum. Since then the phenomenon of 
persistent current in mesoscopic systems has been
discussed quite extensively in the literature both theoretically~\cite{
land,cheu1,cheu2,mont,alts,von,schm,yu,abra,bouz,giam,krav,burme,pichard} 
as well as 
experimentally~\cite{levy,maily,jari,deb,keyser}. 
However, till now the observed experimental features of the persistent  
currents are not well-understood theoretically. A typical example of such
discrepancy between theory and experiment is that the amplitude of the 
measured persistent currents are orders of magnitude larger than the
theoretical predictions. It is believed that the electron-electron
correlation and disorder have major role on the enhancement of  persistent 
currents, but no consensus has yet been reached. Another controversial
issue is that experimentally both
$\phi_0$ and $\phi_0/2$ periodicity has been observed, and it is also found
that
the $\phi_0/2$ oscillations near zero magnetic field exhibit diamagnetic
response. The explanation of these results in terms of the ensemble
averaged persistent currents is also quite intriguing, and the 
calculations show that the disorder averaged current crucially depends
on the choice of the ensemble~\cite{cheu2,mont}. In order to reveal the role of
disorder and electron correlation on the persistent currents,
in this work we focus attention on certain systems which
closely resemble the disorder systems where we do not require any
configuration averaging. These are chemically modulated structures possessing
well-defined long-range order, and, as specific examples we consider
the aperiodic and ordered binary alloy rings.
We confine ourselves to small $1$D rings where 
persistent currents can be calculated exactly, and we obtain many interesting
new results as a consequence of electron correlation and disorder.
One such result is the enhancement of persistent currents in these systems 
due to electron correlation. Another important observation is the 
appearance of 
kink-like structures in the persistent current due to electron-electron
correlation, which may be considered as anomalous Aharonov-Bohm oscillations.
With the recent advancements in sub-micron technology, 
such systems can
be easily fabricated in the laboratory, and in fact, in a recent experiment
Keyser {\em et al.}~\cite{keyser} reported similar anomalous Aharonov-Bohm 
oscillations
from the transport measurements on small rings with less than ten electrons.
Our study may also be helpful to understand the physical properties of 
Benzene-like rings, and other aromatic compounds in the presence of 
magnetic flux.

This paper is organized as follows. In section 2, we present the calculation
of persistent currents in ordered binary alloy rings and investigate their
behavior in the presence of Coulomb repulsion.
In section 3, we describe our results for the incommensurate rings in the
presence of electron-electron interaction. Lastly, we conclude in section 4. 
\section{Ordered binary alloy rings}
 In this section we describe the current-flux characteristics for the
ordered binary alloy rings at $T=0$.
We use the tight-binding Hubbard Hamiltonian with a pure Aharonov-Bohm
flux $\phi$ (in units of $\phi_0$) without any Zeeman term. The magnetic 
vector potential
modifies the hopping integral by a phase factor and the Hamiltonian
for a $N$-site ring becomes
\begin{eqnarray}
H &=& t\sum_{\sigma}\sum_{i=1}^{N}
(c_{i,\sigma}^{\dagger}c_{i+1,\sigma} e^{i\Theta}+h.c.)+ U\sum_{i=1}^{N}
n_{i\uparrow}n_{i\downarrow} \nonumber \\
& & + \sum_{\sigma}\sum_{i=1,3,\ldots}^{N-1}(\epsilon_{A}
n_{i,\sigma}+ \epsilon_{B} n_{i+1,\sigma})
\end{eqnarray}
Here $c_{i\sigma}^{\dagger}(c_{i\sigma})$ is the creation (annihilation)
operator and $n_{i\sigma}$ is the number operator for the electron in the
Wanniar state $|i\sigma>$.
\begin{figure}[ht]
{\centering \resizebox*{7.0cm}{5.5cm}{\includegraphics{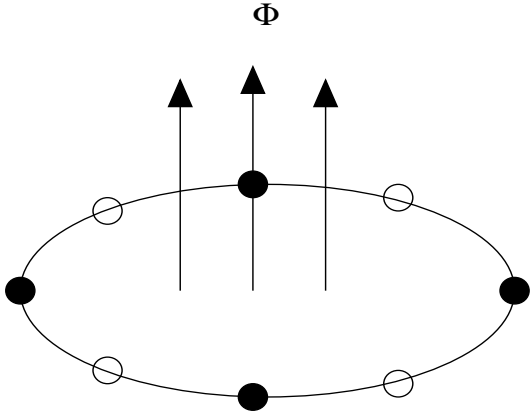}}\par}
\caption{\label{binary}Schematic diagram of a $1$D ordered binary alloy
ring threaded by magnetic flux $\phi$.}
\end{figure}
The parameters $t$ and $U$ are respectively the nearest-neighbor hopping
integral and the strength of Hubbard correlation, and, $\epsilon_A$ and 
$\epsilon_B$ are the site potentials for the $A$ and $B$ type atoms. 
The phase factor is $\Theta=2\pi \phi/N$. Henceforth, 
we take $t=-1$ and use the units $c=e=h=1$. We always choose $N$
to be even so as to preserve the perfect binary ordering of the two types 
of atoms in the ring (see Fig.~\ref{binary}). 

 At zero temperature, persistent current in an isolated ring threaded by  
magnetic flux $\phi$ is given by~\cite{cheu1}
\begin{equation}
I(\phi)=-\frac{\partial E_{0}(\phi)}{\partial \phi},
\end{equation}
where $E_{0}(\phi)$ is the ground state energy. We calculate $I(\phi)$ exactly
by numerical diagonalization of the Hamiltonian.

 Let us now study the behavior of persistent currents in the ordered 
binary alloy rings, and investigate the role of electron-electron interaction
on the currents. In a pure ring consisting of either $A$ or $B$ type
\begin{figure}[ht]
{\centering \resizebox*{8.5cm}{9.5cm}{\includegraphics{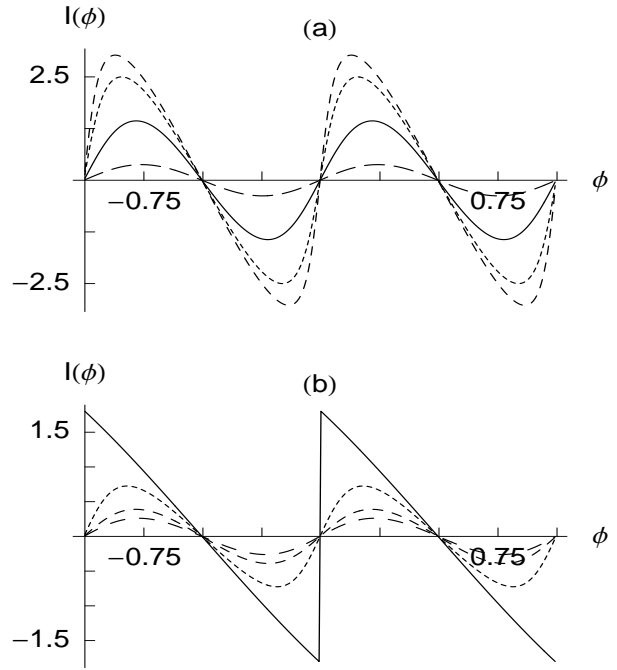}}\par}
\caption{The $I-\phi$ curves of four 
($\uparrow,\uparrow,\downarrow,\downarrow$) electron ordered binary alloy 
rings. (a) $N=4,~ N_e=4$.
The solid, dotted, small dashed and dashed lines correspond to $U=0,~2,~4$
and $10$ respectively, and (b) $N=8,~ N_e=4$. The solid, dotted, small dashed
and dashed lines are respectively for $U=0,~2,~4$ and $6$.}
\label{order}
\end{figure}
atoms without electron correlation, the persistent current as a function of 
$\phi$ is discontinuous at certain points due to ground state degeneracy, and,
the $I-\phi$ curve exhibits a saw-tooth like behavior[3]. In a 
binary alloy ring with $U=0$, this discontinuity completely disappears
as illustrated in Fig.~\ref{order}(a) by the solid curve. The reason for this is
that the binary alloy configuration may be considered as a perturbation
over the perfect ring which lifts the ground state degeneracy, and 
consequently, makes $I(\phi)$ a continuous function of $\phi$. As we
switch on the electron-electron interaction in a pure ring, persistent 
current always decreases with the increase of $U$.
However, in the ordered binary alloy rings, depending on the 
number of electrons $N_e$, we observe enhancement of the persistent current  
for low values of $U$, but eventually it decreases when $U$
becomes very large. Such a situation is depicted in Fig.~\ref{order}(a) where
we display the $I-\phi$ curve for a half-filled ordered binary alloy ring with
four electrons (two up and two down spin electrons). Here we see that for 
$U=2$ (dotted line) and $U=4$
(small dashed line), the current amplitudes are significantly larger 
than the non-interacting case, whereas for $U=10$ (dashed line) the
current amplitude is less than that from the $U=0$ case. This enhancement 
takes place above quarter-filling, i.e., when $N_e>N/2$,
and can be easily understood as follows. As $N$ is even, there are 
exactly $N/2$ number of sites with the lower site potential energy. 
If we do not take into account the electron-electron interaction then
above quarter-filling, it is preferred that some of these lower energy sites 
will be doubly occupied in the ground state. 
As we switch on the Hubbard correlation, 
the two electrons that are on the same site repel each other 
and thus causes enhancement of persistent current. But for large enough $U$, 
the electron-hopping is strongly suppressed by interaction, 
and the current gets reduced. On the other hand, at quarter-filling and
also below quarter-filling, no lower energy site will be doubly occupied 
in the ground state and hence there is no possibility 
of enhancement of persistent current due to Coulomb repulsion. In these
systems we always get suppression
of persistent current with the increase of interaction strength $U$. 
In Fig.~\ref{order}(b), we plot the $I-\phi$ curves for a quarter-filled 
binary-alloy ring with $U=0,~2,~4$ and $6$, and it clearly shows that the
persistent currents are always suppressed by interaction.

\begin{figure}[ht]
{\centering \resizebox*{8.0cm}{8.8cm}{\includegraphics{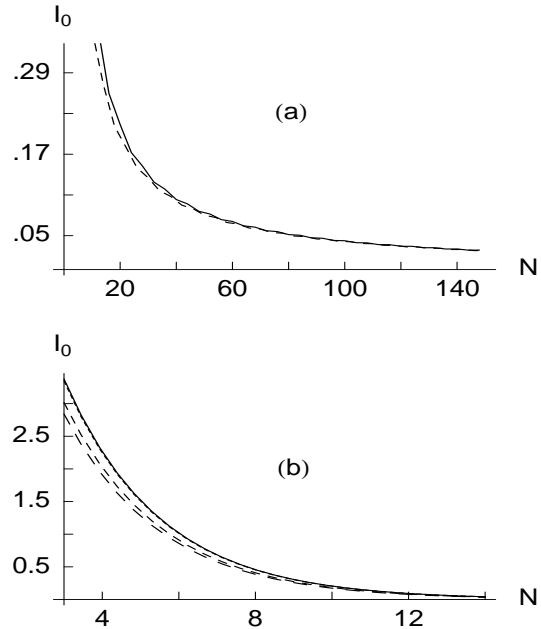}}\par}
\caption{Current amplitude $I_0$ as a function of system size $N$ 
in ordered binary-alloy rings. (a) The solid and dashed lines correspond
the non-interacting rings of size $N=4N_e$ and $N=4N_e+2$ respectively. 
(b) The solid, dotted, small dashed and dashed lines correspond the 
interacting rings of size $N=2N_e$ with $U=0.2,~0.6,~1.5$ and $3$ 
respectively.} 
\label{size1}
\end{figure}

 Let us now describe the behavior of persistent current with system size
$N$ in the ordered binary-alloy rings keeping $N_e/N$ constant. 
For this purpose we calculate current amplitude $I_0$ at some typical 
value of magnetic flux $\phi=0.25$ and in Fig.~\ref{size1} we plot the 
$I_0$ versus $N$ curves. The results for the non-interacting rings are
presented in Fig.~\ref{size1}(a), where the solid and dashed lines
correspond the rings of size $N=4N_e$ and $N=4N_e+2$ respectively.  
Since the dimension of the Hamiltonian matrices for the interacting 
systems increases very sharply with $N$ for higher number of electrons
$N_e$ and also the computational operations are so time consuming, we
present the variations for the interacting rings with size $N=2N_e$ only.
In Fig.~\ref{size1}(b) the results for the interacting rings 
are plotted. The solid and dotted lines, representing the results respectively
for the rings with $U=0.2$ and $U=0.6$, almost coincide with each
other, while, those results for the rings with $U=1.5$ and 
$U=3$ are respectively represented by the small dashed and dashed lines.
These results can predict the variations of current amplitude 
for the rings with size $N=4N_e$ and also $N=4N_e+2$. It is apparent
from Fig.~\ref{size1} that the current amplitude gradually decreases
with system size i.e., we get a converging behavior of current amplitude
with $N$ and most interestingly we see that in the interacting rings
current amplitude converges to zero for any non-zero value of $U$. These
results predict that in a realistic bulk system $I_0$ goes to zero as
soon as the interaction is turned on.
\section{Rings with incommensurate site potentials} 
In this section, we study persistent currents in $1$D rings with quasi-periodic
site potentials, and investigate the effects of electron-electron interaction
on the currents. We describe a $N$-site ring with incommensurate site 
potentials by the Hamiltonian
\begin{eqnarray}
 H= t\sum_{\sigma}\sum_{i=1}^{N}
(c_{i,\sigma}^{\dagger}c_{i+1,\sigma} e^{i\Theta}+h.c.) \nonumber \\
 +\sum_{\sigma}\sum_{i=1}^{N}\epsilon \cos(i \lambda \pi)
c_{i,\sigma}^{\dagger} c_{i,\sigma}+ U\sum_{i=1}^{N}
n_{i\uparrow}n_{i\downarrow}
\end{eqnarray}
where $\lambda$ is an irrational number, and as a typical example we take it as 
the golden mean $\left(\frac{1+\sqrt{5}}{2}\right)$. Setting 
$\lambda=0$ we get back the pure ring with identical site potential $\epsilon$. 
 
\begin{figure}[ht]
{\centering \resizebox*{8.3cm}{5.8cm}{\includegraphics{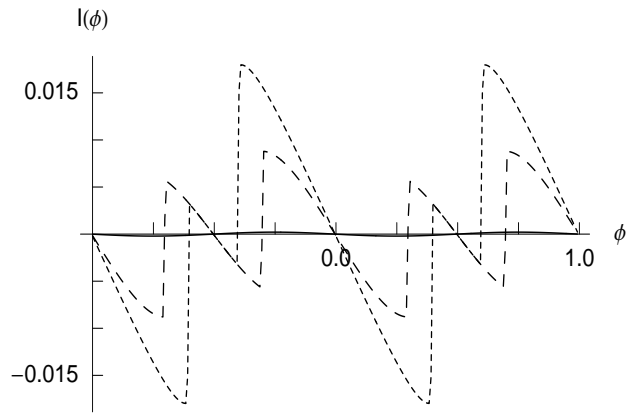}}\par}
\caption{The $I-\phi$ curves for two $(\uparrow, \downarrow)$ electron
incommensurate rings with $N=30$. The solid, dotted and dashed lines 
are respectively for $U=0, 1$, and $3$.}
\label{two}
\end{figure}

 To understand the precise role of electron-electron interaction on 
persistent current in the presence of incommensurate site potentials, 
let us first consider the 
simplest possible system which is the case of a ring with two opposite 
spin electrons. In Fig.~\ref{two} we plot the $I-\phi$ curves for a $30$-site
incommensurate ring with $U=0$ (solid line), $U=1$ (dotted line), and
$U=3$ (dashed line). 
In the absence of any interaction, persistent  
current is greatly reduced by the incommensurate site potential, and in 
fact, Fig.~\ref{two} shows that the $I-\phi$ curve for the non-interacting case
(solid line) almost coincides with the abscissa.
This result can
be easily understood from the argument that in the presence of
aperiodic site potentials, the electronic eigenstates are critical~\cite{
kohmoto,chakra} which
tends to localize the electrons, and thus reduces the current. But this
situation changes quite dramatically as we switch on the electron-electron
interaction. Fig.~\ref{two} clearly shows that electron correlation  
considerably enhances persistent current for low values of $U$. 
This is because the repulsive Coulomb interaction does not favor double 
occupancy of the sites in the ground state, and also it opposes 
confinement of the electrons due to localization. Thus the mobility of
the electrons increases as we introduce Hubbard correlation and gives
enhancement of persistent current. But this enhancement ceases to
occur after certain value of $U$ due to the ring geometry, and 
persistent current then decreases as we increase $U$ further. We also
observe that some strange kink-like structures appear in the
$I-\phi$ characteristics around $\phi=\pm 0.5$ for non-zero values of $U$. 
Quite surprisingly we notice that the persistent currents inside these kinks
are independent of the strength of Hubbard correlation $U$. Let us now
analyze this result. For two opposite spin electrons, total spin $S$
can have the values $S=0$ and $1$. The Hamiltonian of this system for
any $\phi$ can be block diagonalized by proper choice of the basis, and
this can be achieved by taking 
all the basis states in one sub-space with $S=0$, while those in 
the other sub-space with $S=1$. It is easy to see that 
in the sub-space spanned by the basis set with $S=1$, the block Hamiltonian 
is free from $U$, and,
the corresponding energy eigenvalues and eigenstates are $U$-independent.
In the absence of interaction, these $U$-independent energy levels are
always above the
ground level for any $\phi$. But for non-zero values of $U$, one of these
$U$-independent energy levels becomes the ground state energy of the system
in certain domains
of $\phi$. In these regions we have kinks in the $I-\phi$ curves, and
it is obvious that the persistent currents inside these kinks are
independent of the Hubbard correlation $U$. We observe that 
interaction does not alter the $\phi_0$ periodicity of persistent current 
in this two-electron system.

\begin{figure}[ht]
{\centering \resizebox*{8.4cm}{9.5cm}{\includegraphics{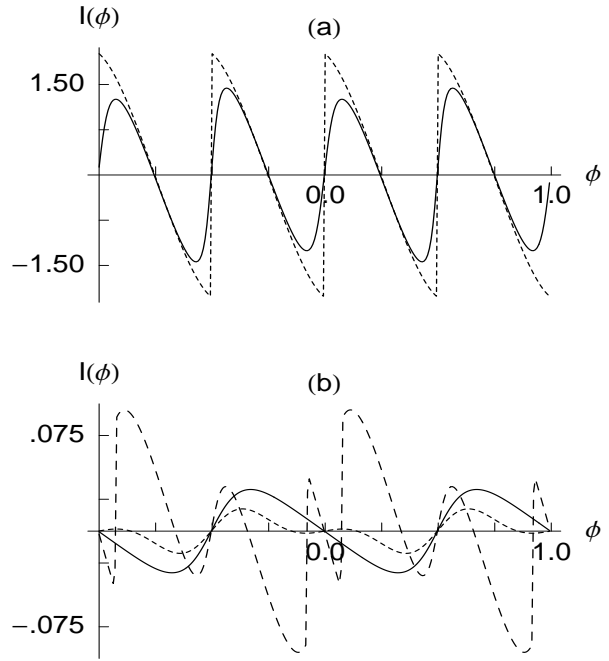}}\par}
\caption{The $I-\phi$ curves for three $(\uparrow, \uparrow, \downarrow)$ 
electron incommensurate rings. (a) Half-filled ($N=3$) systems with $U=4$.
The dotted and solid lines corresponds to $\lambda=0$ and $\lambda \neq 0$ 
respectively. (b) Non-half-filled ($N=12$) systems with $\lambda \neq 0$. 
The solid, dotted and dashed lines are respectively for $U=0,~ 4$, and $50$.}
\label{three}
\end{figure}

 Next we consider incommensurate rings with two up and one down spin 
electrons as the 
representative examples of three-electron systems. The $I-\phi$ 
characteristics for the
half-filled ({\em i.e.}, $N=3, N_e=3$) system with $U=4$ are shown in
Fig.~\ref{three}(a). In this figure, the dotted
line corresponds to a pure ring which exhibits discontinuous
jumps at $\phi=0$, $\pm 0.5$ due to the crossing of the energy levels.
Interestingly, we observe that this pure ($\lambda=0$) half-filled
three-electron system exhibits a perfect $\phi_0/2$ periodicity, and we will
see that this is a characteristic feature of the pure half-filled systems
with odd number of electrons. 
It is evident from the solid curve of Fig.~\ref{three}(a) that this
$\phi_0/2$ periodicity no longer exists as we introduce the incommensurate
site potentials, and we have the usual $\phi_0$ 
periodicity. 
Moreover, in this case $I(\phi)$ becomes a continuous function of $\phi$
as the perturbation due to disorder lifts the ground state
degeneracy at the crossing points of the energy levels.
The characteristic features
of the persistent currents are quite different in the non-half-filled
rings with two up and one down spin electrons, and in Fig.~\ref{three}(b) we 
present the results for a $12$-site ring with incommensurate
site potentials. The solid line is the $I-\phi$ curve for the 
non-interacting electrons ($U=0$), while the dotted and dashed lines 
represent the $I-\phi$ curves for $U=4$ and $U=50$ respectively. The
role of electron-electron interaction on persistent current in the 
presence of incommensurate site potentials 
becomes evident from these curves. For low values of $U$, the $I-\phi$
curve resembles to that for the non-interacting case, and the persistent 
currents
do not show any discontinuity. But for large enough $U$, kink-like 
structures appear in the $I-\phi$ characteristics as illustrated in 
Fig.~\ref{three}(b) by
the dashed line, e.g., around the point $\phi=0$. In this case also, the
kinks are
due to the $U$-independent eigenstates like the two electron systems,
and as explained earlier, the currents inside the kinks are independent
of the strength of correlation $U$. It is 
observed that the persistent currents always have the $\phi_0$ periodicity
in the non-half-filled systems.
We also notice that for the half-filled systems, persistent currents 
always decrease as we increase $U$, while in the non-half-filled rings
we obtain significant enhancement of the currents due to interplay between 
electron correlation and the incommensurate site potentials. 

 As a systematic approach, next we investigate the behavior of persistent
currents in four-electron systems with incommensurate site potentials,
and as the representative examples we consider rings with two up and two
\begin{figure}[ht]
{\centering \resizebox*{8.4cm}{9.5cm}{\includegraphics{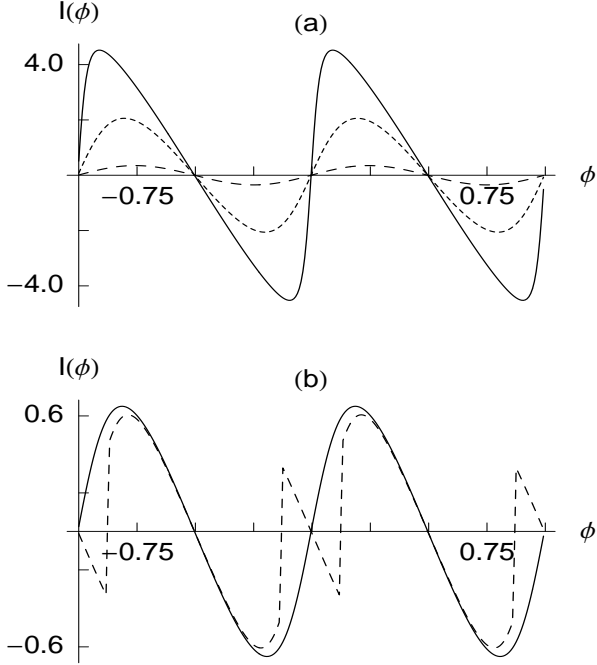}}\par}
\caption{The $I-\phi$ curves for four $(\uparrow, \uparrow, \downarrow,
\downarrow)$ incommensurate rings. (a) Half-filled ($N=4$) systems. 
The solid, dotted and dashed lines corresponds $U=0, 4$ and $10$ 
respectively. (b) Non-half-filled ($N=8$) systems. The solid and
dashed lines are respectively for $U=4$ and $16$.}
\label{four}
\end{figure}
down spin electrons. The $I-\phi$ curves for the half-filled systems are 
plotted in the Fig.~\ref{four}(a). 
The solid, dotted and dashed lines are for
the cases with $U=0$, $U=4$ and $U=10$ respectively. It is evident from 
these curves 
that the current amplitude gradually decreases with the increase of 
interaction strength $U$. This indicates that in the half-filled systems, 
the mobility of the electrons gradually
decreases with the increase of $U$, and we see that for large enough $U$, 
the half-filled system goes to an insulating phase. This kind of behavior 
holds true in any half-filled system because at large enough $U$, 
every site will be occupied by a single electron and
the hopping of the electrons will not be favored due to strong 
electron-electron repulsion. 
In Fig.~\ref{four}(b), we display the $I-\phi$ curves for the non-half-filled
four-electron systems with the aperiodic Harper potential. 
The solid and dotted lines are the $I-\phi$ curves for a $8$-site ring with
$U=4$ and $U=10$ respectively. This figure depicts that for low $U$
persistent current $I(\phi)$ has no discontinuity, but kinks appear in the
$I-\phi$ curve at large $U$. These kinks arise at sufficiently large $U$ due
to additional crossing of the ground state energy levels as we vary $\phi$.
It may be noted that in the present case kinks are due to the
$U$-dependent states, and not from the $U$-independent states as in the 
previous cases. Both for the half-filled or non-half-filled incommensurate
rings with four electrons, we observe that the persistent
currents always exhibit $\phi_0$ periodicity.
\begin{figure}[ht]
{\centering \resizebox*{8.4cm}{5.8cm}{\includegraphics{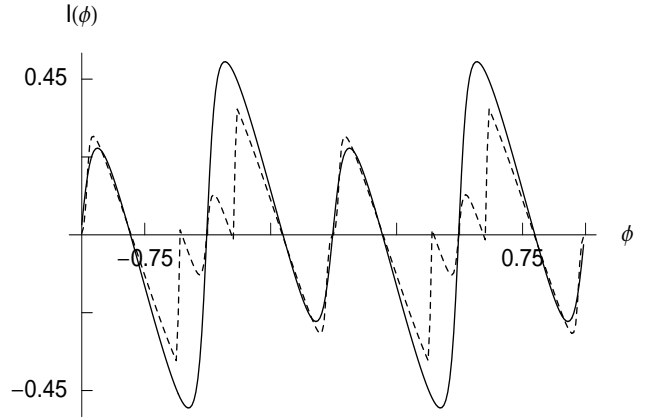}}\par}
\caption{The $I-\phi$ curves for five $(\uparrow, \uparrow, \uparrow,
\downarrow, \downarrow)$ electron incommensurate rings. The curves are for 
the non-half-filled ($N=7$) systems. The solid and dotted
 lines are respectively for $U=18$ and $120$.}
\label{five}
\end{figure}

 Lastly, we consider five electron aperiodic rings, and calculate  
persistent currents in rings with three up and two
down spin electrons. In the pure half-filled case 
({\em i.e.}, $N=5, N_e=5$ and $\lambda=0$) we get $\phi_0/2$
periodicity in persistent current, and we have already observed such
period halving in other pure half-filled systems with odd number of electrons
({\em e.g.}, $N=3,~N_e=3$ and $\lambda=0$).
Like the three-electron half-filled incommensurate systems, also in this case 
the $\phi_0$ 
periodicity of persistent current is restored once we introduced 
incommensurate site potentials. The $I-\phi$ curves for the non-half-filled 
five-electron systems with $N=7$ are shown in Fig.~\ref{five}. 
The solid and dotted lines are respectively for the cases with $U=18$ 
and $U=120$. As in the non-half-filled three-electron system, here also 
kinks appear in the $I-\phi$ 
curves above some critical value of $U$. Another important observation is 
that for large $U$ ($U=120$), the maximum amplitude of the current remains 
finite. This is quite natural since we are considering the systems with 
$N>N_e$ where some sites are always empty so that electrons can hop to the 
empty site, and thus the system always remains in the conducting phase. 
We also see that
in these non-half-filled five-electron rings persistent currents always
have the $\phi_0$ periodicity.  

\begin{figure}[ht]
{\centering \resizebox*{8.0cm}{9.5cm}{\includegraphics{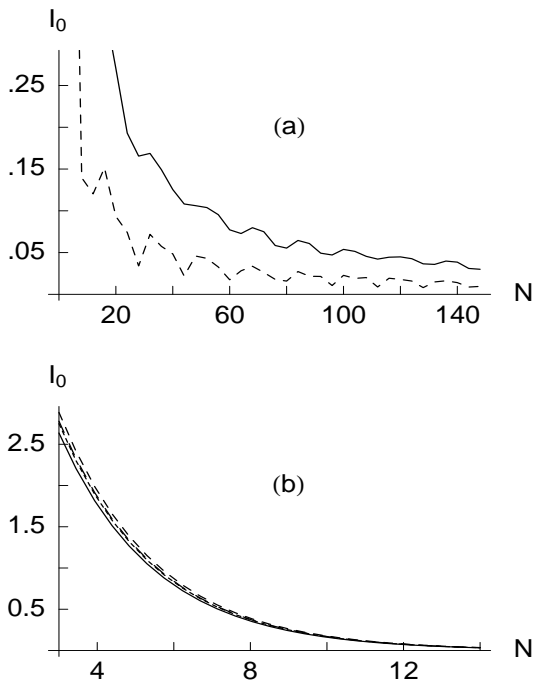}}\par}
\caption{$I_0$ versus $N$ curves for the systems with incommensurate site 
potentials. (a) Non-interacting rings, where the solid and dashed lines
correspond the rings of size $N=2N_e$ and $N=4N_e$ respectively. 
(b) The solid, dotted, small dashed and dashed lines correspond the interacting
rings of size $N=2N_e$ with $U=0.2,~0.6,~1.5$ and $6$ respectively.} 
\label{size2}
\end{figure}
 Now we address the behavior of persistent current with system size $N$
keeping $N_e/N$ constant, for the rings with incommensurate site potentials. 
Like ordered binary-alloy rings here we determine the current amplitude 
$I_0$ at some typical value of $\phi=0.25$ and Fig.~\ref{size2} displays 
the $I_0$ versus $N$ curves. The solid and dashed lines in Fig.~\ref{size2}(a)
respectively correspond the results for the non-interacting rings of size
$N=2N_e$ and $N=4N_e$. On the other hand the variations of the interacting
rings with size $N=2N_e$ are shown in Fig.~\ref{size2}(b). In this figure 
the solid, dotted, small dashed and dashed lines correspond the results
for the rings with $U=0.2,~0.6,~1.5$ and $U=6$ respectively.
These curves reveal that the current amplitude gradually 
decreases with system size $N$ and for large value of $N$ it eventually 
drops to zero value. Thus here we also get a
converging nature of current with the size of the systems and we can say 
that the current amplitude $I_0$ converges to zero for any non-zero value of 
$U$ for large $N$. Thus these results also predict that in a realistic 
bulk system $I_0$ vanishes as soon as the interaction is switched on.
\section{Conclusions}
 In conclusion, we have studied exactly the characteristic features 
of persistent currents in aperiodic and ordered binary alloy rings in
the presence of electron-electron interaction. These systems, which
are neither pure nor disordered rings, have well-defined structural
order, and, our study yields many interesting results due to interplay
between electron-electron interaction and disorder in these systems.
Our main results are : $i)$ In the absence of interaction, the
discontinuity in $I(\phi)$ as a function of $\phi$ disappears due
to disorder. This has been observed both in the ordered binary alloy
rings and also in the aperiodic rings. $ii)$ In pure rings with
electron correlation, we observe both $\phi_0$ and $\phi_0/2$ 
periodicities in
the persistent currents. However, in the incommensurate and ordered
binary alloy rings persistent currents always have the $\phi_0$
periodicity. $iii)$ In the ordered binary alloy rings, above
quarter-filling we get enhancement of persistent current 
for small values of $U$, and the current eventually decreases when $U$
becomes large. On the other hand, below quarter-filling and also at
quarter-filling, persistent current always decreases with the
increase of $U$. $iv)$ An important finding is the appearance of kink-like
structures in the $I-\phi$ curves of the incommensurate rings
only when we take into account electron-electron interaction. Quite 
surprisingly we
observe that in some cases the persistent currents inside the kinks
are independent of the strength of the interaction. These kinks
give rise to anomalous Aharonov-Bohm oscillations in the persistent
current, and recently Keyser {\em et al.}~\cite{keyser} experimentally observed
similar anomalous Aharonov-Bohm oscillations in the transport
measurements on small rings. $v)$ The current amplitude gradually decreases
with $N$ both for the non-interacting and interacting rings, keeping $N_e/N$ 
constant, i.e., we get a converging behavior of 
current with the size of the rings. Most interestingly, we predict that in the
realistic bulk systems $I_0$ vanishes as soon as the interaction is switched
on. Thus the current amplitude vanishes for a ring of macroscopic size.

\end{document}